\documentclass[structabstract]{aa}  
\usepackage{graphicx}
\usepackage{txfonts}
\begin{document}
   \title{Supersonic Cloud Collision - I}

   \author{S. Anathpindika
          \inst{1,2}
          }

   \institute{${}^{[1]}\textrm{S}$chool Of Astronomy \& Physics, Cardiff University, 5-The Parade, UK\\
${}^{[2]}\textrm{I}$ndian Institute Of Astrophysics, 2$^{\textrm{nd}}$ Block - Koramangala, Bangalore 560 034; India\\
              \email{${}^{1}\textrm{S}$umedh.Anathpindika@astro.cf.ac.uk\\
                     ${}^{2}\textrm{s}$umedh\_a@iiap.res.in}
             }

   \date{Received September 00, 0000; accepted March 00, 0000}

 
  \abstract
   {It has long been suggested that shocks might play an important role in altering the form of the interstellar medium (ISM). Shocks enhance gas density and sufficiently dense regions may become self gravitating. Potential star forming clouds within larger molecular clouds, move randomly at supersonic  speeds. }
   {Depending on the precollision velocity, colliding molecular clouds produce a slab that is either shock compressed or pressure confined.  In a sequel of two papers (I \& II), we simulate molecular cloud collision and investigate the dynamical evolution of such slabs. Shocked slabs are susceptible to hydrodynamic instabilities and  in the present paper (I) we study the effect of strong shear between slab layers on the dynamic evolution of a shock compressed gas slab. Both, head-on and off-centre cloud collisions have been examined in this work. We include self gravity in all our simulations.}
   {Simulations presented here, are performed using the smoothed particle hydrodynamics (SPH) numerical scheme. Individual, pre-collision clouds are modelled as pressure confined Bonnor-Ebert spheres. However, for simplicity the  thermodynamic details of the problem are simplified and the gas temperature is simply evolved by a barytropic equation of state. Obviously, the gas, to some extent suffers from thermal inertial effects. However, we note that the dynamical timescale is much smaller than the local sound crossing time so that such effects should have minimum influence.}
   {Highly supersonic cloud collision produces a cold, roughly isothermal shock compressed gas slab. We find that the shocked slab is susceptible to dynamical  instabilities like the gravitational instability, Kelvin-Helmholtz (KH) instability and the non-linear thin shell instability (NTSI). Growth of instabilities within the slab produces structure in it. The NTSI competes with the gravitational instability and the fate of the shocked slab apparently, depends on the relative dominance of either of the two instabilities. Dominance of the NTSI causes turbulent mixing between slab layers and dissipates internal energy.  Eventually the slab collapses to form a thin elongated body, aligned with the collision axis and star formation may commence in it. Our hydrodynamical models discussed here suggest that, high velocity cloud collisions may be a viable mechanism for the formation of observed filamentary structure in the ISM.}
   {}


   \maketitle
%

\section{Introduction}

Spiral shock waves sweep up matter in galactic arms and produce large scale structure and dust lanes  (Roberts 1969; Fleck 1992; Heitsch \emph{et al} 2006; Elmegreen 2007). Often MCs themselves are filamentary and such clouds, and may form out of gravitationally unstable  gas sheets (Nagai, Inutsuka \& Miyama 1998). Detailed studies of molecular line profiles of MCs have revealed that matter within them is highly unevenly distributed and gives them a clumpy appearance, which has led some workers to suggest a fractal model of MCs (Elmegreen 1997). 

 Schneider \& Elmegreen (1977) performed one of the first studies of large scale filamentary structures in the local neighbourhood. Now with improved technology it is possible to map MCs in greater detail. For instance, dense filamentary structures have been observed in the Perseus region where there is evidence for ongoing star formation (Hatchell \emph{et al} 2005). The Orion Integral Filament (OIF) in the Orion A region has been studied in great detail and a number of Young Stellar Objects (YSOs) have been reported to be embedded in it (Chini \emph{et al} 1997; Johnstone \& Bally 1999).

Structure within MCs  appears to be hierarchical and the densest regions are gravitationally bound, see for instance Larson (1981) and Elmegreen (2007). Elongated molecular and atomic structures are common occurrences in the ISM and a number of possible formation mechanisms have been suggested (c.f. Kluging \& Heiles 1987). One possible mechanism is the dynamical interaction between fluid flows within MCs (Klessen \& Burkert 2000; McLow \& Klessen 2004). Turbulence decays rather quickly, on a timescale equal to the dynamical period of evolution of the cloud. The importance of turbulence in star formation has been highlighted by several workers (e.g. Sasao 1973; Klessen, Heitsch \& McLow 2000; Klessen \& Burkert 2001 and references therein). Gas slabs resulting from collision of stellar winds from binary systems expanding at supersonic speeds become thermally unstable due to rapid cooling, also turbulent motion within shells makes them unstable to shearing instabilities (Stevens, Blondin \& Pollock 1992). Similarly, powerful ionising radiation emitted by young OB associations or supernova remnants  sweeps up gas and forms a dense shell that is gravitationally and thermally unstable (Strickland \& Blondin 1995; Dale, Clarke \& Bonnell 2007). Low velocity MC collision produces gravitationally unstable gas slabs (Bhattal \emph{et al} 1998). 

 Analogous to random motion of smaller clouds, larger MCs (Giant Molecular clouds, GMCs) in galactic arms also exhibit random motion. Line width of CO emission from GMCs suggests that clouds have a random velocity dispersion typically of order a few km s$^{-1}$, superposed on their systematic galactic motion (Spitzer 1968; Solomon \emph{et al} 1987). GMCs are largely supported by internal turbulence and have lifetimes of about a few tens of Myrs, before being dispersed due to stellar feedback (e.g. McLow 1999; Padoan \& Nordlund 1999; Klessen, Heitsch \& McLow 2000; Blitz \emph{et al} 2007).

Smaller clouds within GMCs themselves have turbulent velocities of $\sim$1-2 km s$^{-1}$ on a parsec scale (e.g. Larson 1981; Elmegreen 2000). Collision between such clouds, moving randomly with supersonic speeds is an efficient mode of dissipating kinetic energy. High velocity cloud collision is a violent phenomenon that might trigger star formation in the resulting  composite object. The gas bodies resulting from dynamically interacting fluid masses already contain the necessary seeds of dynamic instability.  Magnetic field threading the GMCs is generally weak, of order a few $\mu$G only, and does not seem to play any significant role in damping turbulence within GMCs. However, the magnetic field strength could increase dramatically  due to post-shock enhancement of gas density, which in turn may cause further compression of the shocked gas (e.g. Hartmann, Ballestros-Paredes \& Bergin 2001).

The problem of colliding clouds has been  studied numerically in the past for instance, by Stone (1970 a,b), Hausman (1981), Lattanzio \emph{et al} (1985) and Hunter \emph{et al} (1986); more recently by Bhattal \emph{et al} (1998) and Klein \& Woods (1998), among others. Stone (1970 a) studied this problem in one dimension by simulating MCs as planes having infinite spatial extent and found that post-collision, the clouds coalesced and the resulting gas slab then expanded. The same experiment, in two dimensions also yielded a similar result (Stone 1970b). Hausman (1981) performed particle simulations of head-on and off-centre cloud collisions. However, the results were corrupted due to particle penetration during collision.

Later Hunter \emph{et al}, in their grid simulation of colliding flows found that such collisions produced slabs which according to them were unstable to  Rayleigh-Taylor like instability, but could not establish the exact nature of this instability. A similar conclusion was drawn by Stevens, Blondin \& Pollock (1995) and Klein \& Woods (1998). However, both found evidence for non-linear growth of instabilities starting in thin shells within the slab. In fact the latter suggested occurrence of the non-linear thin shell instability (NTSI) in the shocked slab, resulting from cloud collisions.

In this paper we demonstrate the occurrence of this instability in our SPH simulations of highly supersonic (10 km s$^{-1}$ to 20 km s$^{-1}$) head-on cloud collision. For the purpose of this work, we use  clouds with only token masses, a few orders of magnitude smaller than that of real star forming clouds. Although we test one model with slightly massive clouds, even their masses are still modest. Limited computational resources is a major hurdle in using larger number of particles, resolution would be severely compromised in case clouds with realistic masses were used. The precollision velocity of individual clouds however, is representative  of the dispersion velocity of clouds in the galactic plane. The possible effect of the SPH artificial viscosity (AV), on the growth rate of the NTSI is left for a future paper.

The plan of the paper is as follows. In \S 2 we briefly describe the SPH code employed for our cloud collision experiments. In \S 3 we discuss the procedure to set up the initial conditions for cloud collision experiments and briefly discuss the models tested. In \S 4 we discuss the NTSI and simulation results  and conclude in \S 5. Since the column density spans roughly eight orders of magnitude, all column density plots presented here are plotted on a logarithmic scale. The column density is measured in M$_{\odot}$pc$^{-2}$ and the distance in pc.


\section{Smoothed Particle Hydrodynamics (SPH)}

SPH is a Lagrangian interpolation scheme used to treat complicated  problems in astrohydrodynamics. Under the scheme,  the fluid being investigated is treated as an ensemble of a large number of particles or fluid markers. The particles in this case are not point masses in the strictest sense, but have a finite spatial extent  (Gingold \& Monaghan 1977; Lucy 1977).

All the simulations presented here have been performed using the DRAGON SPH code. It is a well tested code, the relevant details of which can be found in (Goodwin, Whitworth \& Ward-Thompson 2004). It uses the Barnes-Hut tree for gravity calculation ($\theta_{crit}\sim 0.45$) and also includes quadrupole moments of remote cells. The forces are integrated using a second order Runge-Kutta integration scheme. Each particle has (50$\pm$5$\%$) neighbours. The code employs a multiple particle time stepping scheme as against  global time stepping and uses the standard SPH artificial viscosity.

Extremely dense agglomerations of particles in the computational domain are replaced by \emph{sinks}. A sink is characterised by its density, $\rho_{sink}$, and radius, $R_{sink}$. The concept of a sink particle was first introduced in SPH by Bate, Bonnell \& Price (1995). Rest of the gas particles in the computational domain interact with the sink only through gravity. SPH particles coming within a predefined radius, called the sink radius ($R_{sink}$), are accreted by it, provided particles being accreted satisfy certain criteria (Bate, Bonnell \& Price 1995). In all the simulations presented here, the sink density is held  fixed at $\rho_{sink}=10^{-12}$g cm$^{-3}$. $R_{sink}$ is so chosen that the initial sink mass, $M_{sink}$, is comparable to the minimum resolvable mass in the simulation, $M_{min}$. Now, 
\begin{displaymath}
M_{min} = N_{neibs}\cdot m_{i}
\end{displaymath}
where, $N_{neibs}$ is the number of nearest neighbours of an SPH particle, $m_{i} = \frac{M_{cld}}{N_{gas}}$ and $M_{sink} = \frac{4}{3}\pi R^{3}_{sink}\rho_{sink}$. In view of the above mentioned requirement, 
\begin{displaymath}
R_{sink} \sim \Big(\frac{N_{neibs}M_{cld}}{4N_{gas}\rho_{sink}}\Big)^{1/3}.
\end{displaymath}
In the present work, the sink particle represents a protostellar core.


\begin{table*}
\caption{List of simulations performed with their relevant physical details.}   
\label{table:1}      
\begin{center}      
\begin{tabular}{c l l r r c }     
\hline\hline       
  Serial & Experimental &  Pre-collision & $P_{ext}$ & Number of & Head-on\\
No. & details & Mach number $(\mathcal{M})$ & & particles &  \\

 \hline
  1 & $M_{cld1}$ = $M_{cld2}$ = 100 M$_{\odot}$ & 35 & Yes & $N_{tot}$ = 34893 & Yes \\
& $R_{cld1}$ = $R_{cld2}$ = 1 pc, $T_{cld}$ = 84 K & & & $N_{gas}$ = 3516  &  \\
\hline
 2 & $M_{cld1}$ = $M_{cld2}$ = 50 M$_{\odot}$ & 25 & Yes & $N_{tot}$ = 129427 & Yes \\
& $R_{cld1}$ = $R_{cld2}$ = 0.8 pc, $T_{cld}$ = 54 K & & & $N_{gas}$ = 16000 &   \\
\hline
 3 & $M_{cld1}$ = $M_{cld2}$ = 400 M$_{\odot}$ & 25 & No & $N_{gas}$ = 120000 & Yes \\
& $R_{cld1}$ = $R_{cld2}$ = 2.0 pc, $T_{cld}$ = 170 K & & & & \\
\hline
 4 & $M_{cld1}$ = $M_{cld2}$ = 50 M$_{\odot}$ & 25 & No & $N_{gas}$ = 40000 & No \\
& $R_{cld1}$ = $R_{cld2}$ = 0.8 pc, $T_{cld}$ = 54 K & & & & $b$ = 0.2 pc\\

\hline                  
\end{tabular}
\end{center}
Note : The numbers on the left hand side below indicate the respective column numbers of the table.\\
(2) Physical details of individual clouds in a simulation. \\
(3) The pre-collision Mach number of individual clouds.\\
(4) Indicates if clouds are confined by external pressure, $P_{ext}$, exerted by the ICM particles. \\ 
(5) Total number of particles ($N_{tot}$) used in a simulation (ICM + GAS). The number of gas particles ($N_{gas}$) are separately mentioned. \\
(6) Indicates if the clouds were allowed to collide head-on/off-centre; $b$ is the impact parameter of collision. \\
\end{table*}

  \section{Numerical experiments:}
\subsection{Setting the initial conditions (ICs)}
   
  We first describe the scheme to assemble particles within a MC. The MC is modelled as an external pressure confined isothermal sphere. An isothermal sphere is characterised by its dimensionless radius, $\xi$, and it has been well established that only those spheres having radii smaller than a critical radius, $\xi_{crit}$ = 6.45, can remain in equilibrium (Chandrasekhar 1939). For our purposes here, we choose $\xi_{B}\equiv\xi$ = 3, which ensures that the sphere is supported against self gravity.

For simplicity we further assume that the sphere contains only molecular hydrogen and if $\bar{m}$ is the mean molecular mass of a gas particle, then the isothermal sound speed $a_{0}$ is defined as 
\begin{equation}
a_{0} = \Big(\frac{k_{B}T_{cld}}{\bar{m}}\Big)^{\frac{1}{2}}.
\end{equation}
Here $k_{B}$ and $T_{cld}$ respectively, are the Boltzmann constant and uniform gas temperature in the sphere.

The mass interior to this sphere, $M(r_{B})$, is 
\begin{equation}
M_{sphere}\equiv M(r_{B}) = 4\pi R^{3}_{0}\rho_{c}\Big(\xi^{2}\frac{d\psi}{d\xi}\Big)_{\xi_{B}},
\end{equation}
where $\psi(\xi)$ is called the Emden function (Chandrasekhar 1939) and the physical radius of the sphere, $r_{B} = R_{0}\xi_{B}$, $R_{0}$ having the dimension of length is a distance scale factor, and defined as
\begin{equation}
R_{0} = \frac{a_{0}}{(4\pi G\rho_{c})^{\frac{1}{2}}}.
\end{equation}
 Taking the ratio of $M_{sphere}$ with $r_{B}$  we get,
\begin{displaymath}
   \frac{M_{sphere}}{r_{B}}=\frac{a_{0}^{2}\mu(\xi_{B})}{G\xi_{B}}
\end{displaymath}
\begin{equation}
\Rightarrow a_{0}^{2}=\frac{GM_{sphere}\xi_{B}}{r_{B}\mu(\xi_{B})},
\end{equation}
where we define
\begin{displaymath}
\mu(\xi_{B}) = \Big(\xi^{2}\frac{d\psi}{d\xi}\Big)_{\xi_{B}}.
\end{displaymath}
The functions $\psi(\xi)$ \& $\mu(\xi)$ are tabulated by integrating  the isothermal Lane - Emden equation. The temperature within the sphere, $T_{cld}$, is calculated using equations (1) and (4).



 Particles within the sphere are randomly assembled using three random number generators, each producing a random number in the range (0,1). A random number $\mathcal{R}_{r}$ is chosen such that 
\begin{displaymath}
\frac{M(r)}{M_{sphere}}=\frac{\mu(\xi)}{\mu(\xi_{B})}\equiv \mathcal{R}_{r}
\end{displaymath}
\begin{equation}
\Rightarrow  \mu(\xi)=\mu(\xi_{B})\mathcal{R}_{r},
\end{equation}
Knowing the RHS of equation (5), the corresponding dimensionless radius $\xi$ can be obtained from the Lane - Emden table. The radial distance $r$ of a particle  within the sphere is then given by $r=\mathcal{R}_{r}\xi$.  The probability that a particle will lie in an infinitesimal interval of the poloidal angle ($\theta$) and ($\theta+d\theta$) is then,
\begin{displaymath}
p_{\theta}d\theta=\frac{\sin\theta}{2}d\theta.
\end{displaymath}
The total probability over the range of the angle $\theta$ is,
\begin{equation}
P(\theta)=\int_{\theta'=0}^{\theta'=\theta} \frac{\sin\theta'}{2}d\theta'=\frac{1-\cos\theta}{2}\equiv \mathcal{R}_{\theta},
\end{equation}
where $\mathcal{R}_{\theta}$ is the second random number. Thus,
\begin{displaymath}
\cos\theta = 1-2\mathcal{R}_{\theta},
\end{displaymath}
and finally the azimuthal angle, $\phi$, of a particle  is fixed as,  $\phi=2\pi \mathcal{R}_{\phi}$; $\mathcal{R}_{\phi}$ being a third random number.  The $(r,\theta,\phi)$ coordinates are then converted to their cartesian form in the usual way.

The density of this sphere is
\begin{equation}
\rho(r_{B}) = \frac{a^{6}_{0}}{4\pi G^{3}M^{2}_{sphere}}\Big(\xi^{2}\frac{d\psi}{d\xi}\Big)_{\xi=\xi_{B}}^{2}\cdot e^{-\psi(\xi_{B})},
\end{equation}
and $\xi_{B} = 3$, by choice. However, such a sphere in isolation will simply diffuse away and therefore  needs to be spatially confined with a finite external pressure, $P_{ext}$, acting on its boundary. The magnitude of this pressure is
\begin{equation}
P_{ext} = \frac{a^{8}_{0}}{4\pi G^{3}M^{2}_{sphere}}\cdot \mu(\xi_{B})e^{-\psi(\xi_{B})},
\end{equation}
where all symbols have their usual meanings (Chandrasekhar 1939). 

We simulate the warm external intercloud medium (ICM) by assembling particles in an envelope of finite thickness ($\delta h_{B}$).  Here $h_{B}$ is the average smoothing length of a single SPH particle and $\delta$ is a non-zero integer defining the thickness of the envelope. Particles in this envelope interact with the gas particles within the cloud only by exerting hydrodynamic pressure of magnitude $P_{ext}$, defined by equation (8) above.

%
   \begin{figure}
   \centering
   \includegraphics[angle=270, width=6cm]{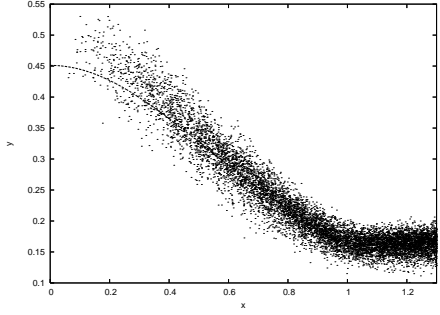}
      \caption{This figure shows the SPH density, $\rho_{SPH}$, of the truncated isothermal sphere ($\xi_{B}$=3). The dashed continuous line is the expected density profile, $\rho(r_{B})$, for this sphere. ($x\equiv\xi,y\equiv\rho_{SPH}$), $N_{gas}$ = 1500.}
   \end{figure}
%

To prevent particles from diffusing across the cloud-ICM interface, care must be taken to avoid a gradient in particle number density. To ensure this, we first calculate the particle number density, $\hat{n}_{B}$, near the cloud edge and then evaluate the number of particles to be assembled in this envelope, $N_{env}$. For the purpose, we first obtain the expression for density at the cloud edge in its dimensionless form, $\hat{\rho}_{B}$. A trivial mathematical manipulation of equation (7) yields,
\begin{equation}
\hat{\rho}_{B}=\frac{\xi_{B}^{3}e^{-\psi(\xi_{B})}}{4\pi \mu(\xi_{B})}.
\end{equation}
The particle  number density at the edge of the sphere is $\hat{n}_{B}=\frac{\hat{\rho}_{B}}{m_{i}} \equiv N_{gas}\hat{\rho}_{B}$, where $m_{i}$, the mass of the $i^{th}$ particle in our test sphere is $m_{i}=\frac{1}{N_{gas}}$ and the volume of the ICM envelope is  $\frac{4\pi [(1+\delta\cdot h_{B})^{3}-1]}{3}$, for a sphere of unit mass. The number of particles to be assembled in the envelope, $N_{env}$, is then the product of this volume with the particle number density, $\hat{n}_{B}$.

 The radial coordinate, $r$, of a particle in the envelope is fixed by choosing a random number, $\mathcal{R}_{r}$, such that the volume interior to this radius is proportional to the volume in the envelope i.e
\begin{displaymath}
\mathcal{R}_{r}= \frac{(r^{3}-1)}{((1+\delta\cdot h_{B})^{3}-1)}.
\end{displaymath}
The polar angles ($\theta,\phi$) of individual particles are determined as before. The polar coordinates of particles are then converted to their cartesian form. With this, the assembly of the sphere-envelope system is completed.

To test this scheme of assembling particles, we assemble 1500 particles within the sphere ($N_{gas}$) and with a choice of $\delta$ = 5, $N_{env}$ = 6159. This assembly is then placed in a periodic box and allowed to evolve for a fraction of the sound crossing time. The periodic box is simply meant for ghosting particles i.e. particles leaving from one face of the cube re-enter from the opposite face. Fig. 1 shows the SPH density of a truncated isothermal sphere, after about half sound crossing time and seems to be in reasonable agreement with the expected density profile, marked by the dashed continuous curve. 
 
  \begin{figure*}
   \centering
\vbox to 90mm{\vfil
   \includegraphics [width=10cm]{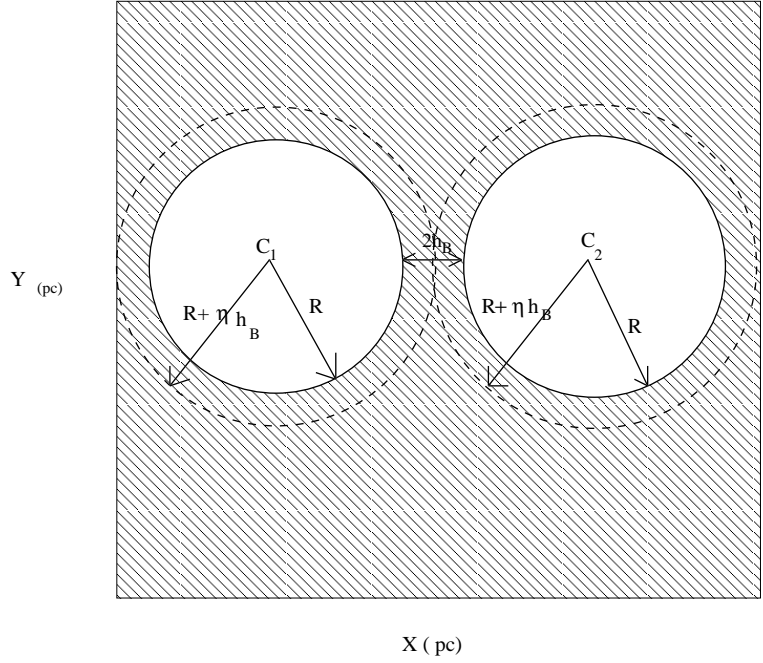}
      \caption{Schematic diagram of the initial set up of colliding clouds. See text above for description.}
              \vfil}
         \label{FigVibStab}
   \end{figure*}

The box of the ICM particles is set up in a manner similar to the envelope of $P_{ext}$ particles for the truncated isothermal sphere, described above. The dimensions of the box are so set that each sphere has an envelope of thickness  $3h_{B}$. Fig. 2 below shows a schematic diagram of the  initial set up. It shows the precollision MCs barely touching each other in the plane of collision. Circles drawn with continuous lines represent the precollision clouds. Each circle is circumscribed by respective dashed circles which show the extent of the ICM envelope for each cloud.

Since we have opted to use physically identical clouds, each one of them has the same mass, $M_{cld}$, radius, $R_{cld}$, and an envelope of the ICM particles of thickness $\delta h_{B}$. The length of the ICM box along the  axis of collision (the $x$ axis in Fig. 2) is 
\begin{displaymath}
x = 4(R + \delta h_{B}),
\end{displaymath}
while that along the $y$ and  $z$ axes respectively,  is
\begin{displaymath}
y = z = 2(R + \delta h_{B}),
\end{displaymath}
and $\delta = 3$ in the present case. A short coming of the SPH code used, is that it requires a cubic periodic box. Each side of the cubic box therefore has length $x$ pc. 

Let $N_{gas1}$ and $N_{gas2}$ be the number of gas particles in each cloud (provided externally). The number of particles to be assembled in the ICM envelope of respective spheres is  $N_{env1}$ and $N_{env2}$. The region marked by stripes in Fig. 2 is that occupied by the ICM particles, the volume of which, can be trivially calculated. We also know the required particle density, $\hat{n}_{B}$, in this region using which the number of particles to be assembled in this region is calculated. The total number of particles in the box is simply the sum of $N_{gas1}$, $N_{gas2}$, $N_{env1}$, $N_{env2}$ and the number of ICM particles in the striped region of the box. 

Particles in this box are assembled on a HCP lattice. Particles within the desired cloud radius are then scooped out and replaced with corresponding gas particles. The second cloud is assembled in a similar manner. This concludes the process of setting up the ICs.

\subsection{Equation of state (EOS) }

In the simulations discussed here, we use a slightly modified barytropic EOS given by,
\begin{displaymath}
\frac{P(\rho)}{\rho} = (k_B/\bar{m})\times
\end{displaymath}
\begin{equation}
\left\{ \begin{array}{ll}
\Big(\frac{T_{cld}}{\textrm{K}}\Big) &; \rho\le10^{-21} \textrm{g cm}^{-3}\\
\Big(\frac{T_{cld}}{\textrm{K}}\Big)\Big(\frac{\rho}{\textrm{g cm}^{-3}}\Big)^{\gamma-1} &; 10^{-21}\textrm{g cm}^{-3} <\rho\le \\
& 5\times 10^{-21}\textrm{g cm}^{-3}\\
\Big(\frac{T_{ps}}{\textrm{K}}\Big) &; 5\times 10^{-21}\textrm{g cm}^{-3} <\rho\le\\&10^{-18}\textrm{g cm}^{-3}\\
\Big(\frac{T_{cld}}{\textrm{K}}\Big)\Big[1 + \Big(\frac{\rho}{10^{-14}\textrm{g cm}^{-3}}\Big)^{\gamma-1}\Big]&; \rho > 10^{-18}\textrm{g cm}^{-3}.
\end{array} \right.
\end{equation}
Here, $T_{cld}$, is the pre-collision cloud temperature, $T_{ps}$,  the post shock temperature, $k_{B}$, the Boltzmann constant, $\bar{m}$, the mean mass of a hydrogen molecule and, $\gamma$, the adiabatic constant which, in the present case is $\frac{5}{3}$. This is due to our assumption of monatomic gas within individual clouds.

The density in the thin dissipative region downstream of a strong adiabatic shock, is at most four times the preshock density, a direct consequence of the Rankine-Hugoniot jump condition. Shocked gas in  astrophysical environs, that can radiate efficiently,  cools and becomes approximately isothermal. In the next few lines we attempt to justify our assumption that the post collision gas slab in the models tested here, is approximately isothermal. That would be a rather straightforward task in case, a post-shock cooling scheme were employed.

Nonetheless, one may summon for the purpose, the average post shock density, temperature and a model for radiative cooling proposed by Semadeni \emph{et al} (2007). The temperature jump condition, for models 2 and 3 implies a post shock temperature, $T_{ps}$, of order 6,000 K and 11,000 K respectively, while that for model 4 is the same, as in model 2. The cooling rate suggested by the latter authors would then correspond to a cooling time, $\tau_{cool}$, of order a few thousand years only. In that case, the cooling length, $l_{cool}$ ($l_{cool}\sim \tau_{cool}a_{0}$), over which the shocked gas downstream a shock cools to the original preshock temperature, would only be a tiny fraction of the observed slab thickness in the simulations performed. A small cooling length and cooling time (compared to the sound crossing time), presage the assumption  of isothermality behind a shock.

To conclude the argument, let us additionally calculate $\tau_{cool}$ using the relation,
\begin{equation}
\tau_{cool}\sim \frac{2.4\times 10^{5}}{\bar{n}_{H}} \ \textrm{yrs}.
\end{equation}
Here $\bar{n}_{H}$ is the post-shock number density of atomic hydrogen (Spitzer 1978). Even by conservative estimates, if we were to assume a compression factor of 4, as in a strong adiabatic shock, the post shock density  would be of order 10$^{3}$ cm$^{-3}$, so that equation (11) implies $\tau_{cool}$ to be about a few hundred years only, much smaller than the local sound crossing time. 

For the models tested here, the post shock density is of order $\sim$10$^{-21}$ g cm$^{-3}$. Thus, for gas density below this threshold, we adopt an isothermal EOS and maintain the gas at the initial cloud temperature, $T_{cld}$. Following the cloud collision, for densities above this threshold and less than 5$\times$10$^{-21}$ g cm$^{-3}$, we adopt an adiabatic EOS to model the shock. The gas is then  held at a post shock temperature, $T_{ps}$. When the shocked gas becomes sufficiently dense ($\sim \mathcal{M}^{2}\rho_{cld}\sim 10^{-18}$ g cm$^{-3}$), its temperature is brought down to that in the initial, pre-collision clouds, $T_{cld}$. The cold, approximately isothermal gas slab is then so maintained up to a density of $10^{-14}$ g cm$^{-3}$. Above this threshold, the EOS flips from being isothermal to adiabatic. This final phase is meant to follow putative star forming clumps in the shocked slab.

\subsection{Models tested}

The cloud collision  experiments presented here have been listed in Table (1) above. The table  provides the relevant physical details of an experiment. We arbitrarily choose the precollision Mach number, $\mathcal{M}$, and the cloud mass, $M_{cld}$. The corresponding cloud radius, $R_{cld}$, is determined using the Larson's scaling relation between  mass and radius,
\begin{equation}
R_{cld}(\textrm{pc}) = 0.1\ (\textrm{pc})\ \Big(\frac{M_{cld}}{\textrm{M}_{\odot}}\Big)^{0.5}
\end{equation}
(Larson 1981).

The fate of the post-collision slab depends largely on the initial conditions. We observe contrasting features in the evolution of the shocked slab in models 1, 2 and 3. While model 1 is under resolved and just a crude test case, the latter are much better resolved, with larger number of gas particles. 

\subsubsection {Model 1} 

In this case the clouds collide head-on at a highly supersonic speed. Post-collision the clouds coalesce to form a shock compressed gas slab, that evolves through the compression and re-expansion phases described below. Artificial viscosity dissipates the kinetic energy of colliding clouds and heats up the shocked slab, whose temperature is denoted as $T_{ps}$.  The gas slab in this case does not become Jeans unstable, presumably due to the following two reasons- (i) insufficient resolution as too few particles have been used and (ii) the post-shock density being less than that required to support the Jeans instability. 

Consequently, the slab simply re-expands along the collision axis. Also, there is no  evidence of any other instability in the shocked slab. This is obviously due to the poor resolution in this model. The fate of such a collision may be envisaged to be a diffuse cloud, a part of the ICM. In order to test this outcome, we perform a similar experiment in the next case.

\begin{figure*}
\vbox to 150mm{\vfil 
\includegraphics[angle=270, width=14.0cm]{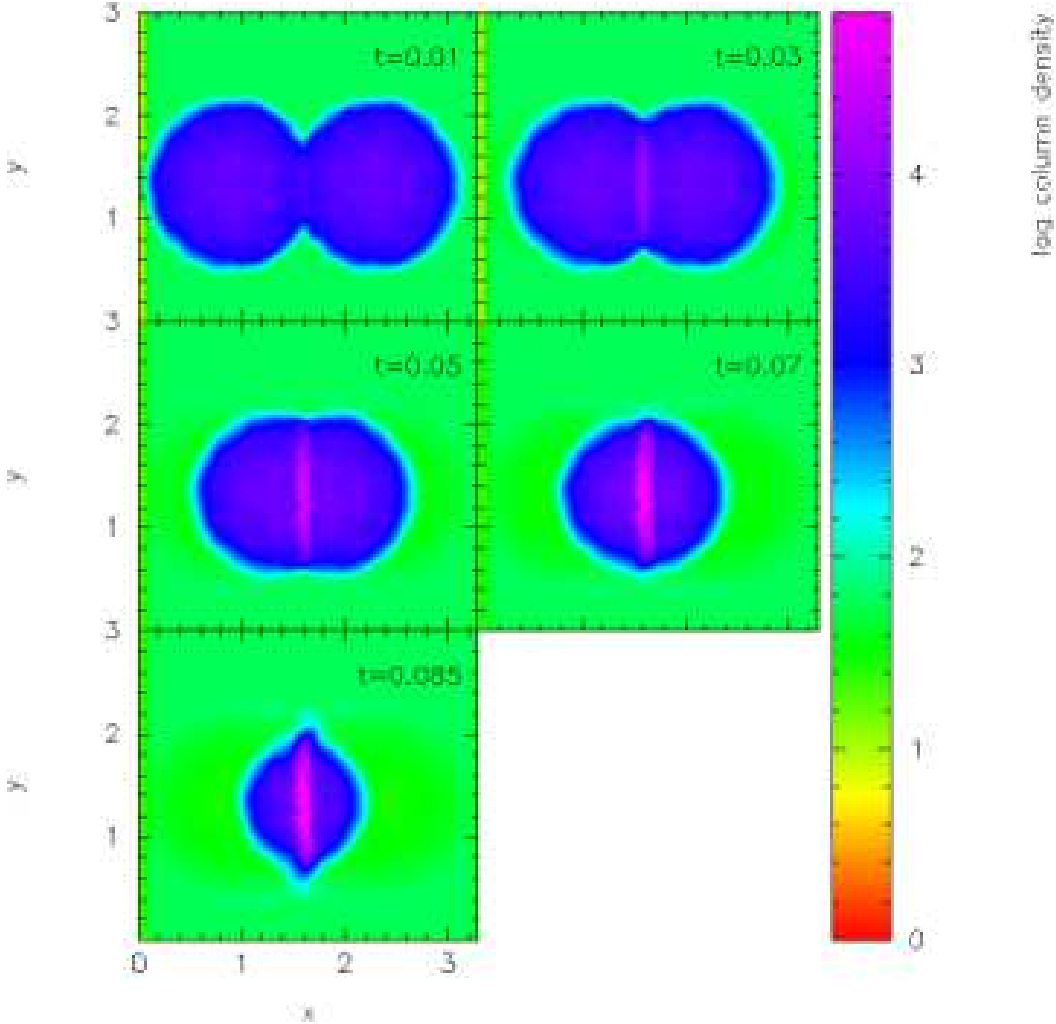}
\caption{A column density plot showing a time (in Myr) sequence of the colliding clouds in model 2. The collision produces a shocked slab that develops bending modes and kinks in the slab are visible in the plots on the second row. Some material also squirts out, \emph{jets}, from the top and bottom ends of the slab (t = 0.85 Myr). } 
\vfil} \label{landfig}
\end{figure*}

\begin{figure}
\centering
\includegraphics[angle=270, width=8.5cm]{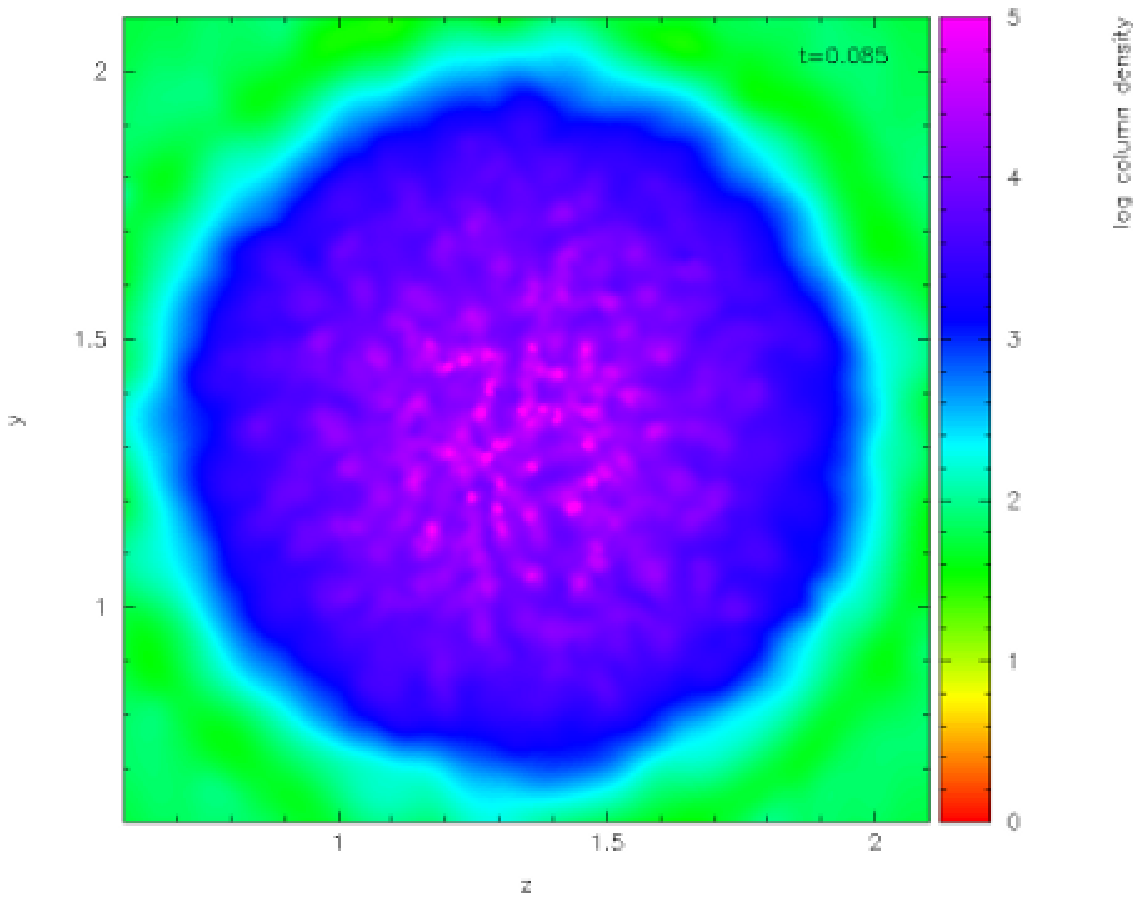}
\caption{A column density plot of the fragmented slab in model 2. We can see a network of filaments with several clumps in them, marked as bright pink blobs. The clumps, actually are sink particles. Time measured in Myr.} \label{landfig}
\end{figure}

\subsubsection{Model 2}

Although the pre-collision cloud speed in this case is some what lower than in the previous case, individual clouds are still highly supersonic. Column density plots in Fig. 3 show a time sequence of the colliding clouds in this model. Formation of the contact layer  can be seen in the picture corresponding to $t$ = 0.01 Myr. The slab then develops a kink on it (the bending of the slab) which then rapidly amplifies, as can be seen in the snapshots corresponding to $t = 0.03$ Myr, $t= 0.07$ Myr and $t = 0.085$ Myr. Although the slab layers are sheared, the gravitational instability dominates in this case. The shocked slab fragments and forms a network of filaments. Naively speaking, the slab develops a number of holes in it and gas ablates in to filaments. 

Subsequently these filaments become gravitationally unstable and produce dense clumps in them. The face-on picture of the fragmented slab in Fig. 4 shows  filamentary network and the clumps within them. The clumps can be  identified as bright pink blobs in the plot. The gravitational instability grows rapidly and the slab fragments on a timescale much smaller than $t_{cr}$, the cloud crushing time defined in \S 4.2 below. The criterion for dominance of the gravitational instability over shearing instabilities is
\begin{equation}
\bar{n} \gtrsim 4.7\times 10^{2}\Big(\frac{\eta}{\textrm{pc}}\Big)\Big(\frac{\lambda_{NTSI}}{\textrm{pc}}\Big)^{-3}\Big(\frac{T}{\textrm{K}}\Big)\ \textrm{cm}^{-3},
\end{equation}
where $\bar{n}$ is the gas number density in the slab, $\eta$ is the amplitude of perturbation and $\lambda_{NTSI}$ is the length of an unstable mode (Heitsch, Hartmann \& Burkert 2008).

 From the simulation we find that $\eta \sim$ 0.02 pc and $\lambda_{NTSI}\sim$ 0.01 pc, so that equation (13) yields $\bar{n} \sim 10^{5}$ cm$^{-3}$. In the present case the post-shock density is $\sim 10^{-18}$ g cm$^{-3}$, so that the corresponding number density is $\sim 10^{6}$ cm$^{-3}$, which is nearly an order of magnitude larger than the calculated $\bar{n}$. The shocked slab is therefore susceptible to gravitational instability. This outcome however, contrasts with that of the previous case where the gas slab evolved in to a diffuse cloud.

\subsubsection{Model 3}

This model is similar to the previous two except that the precollision clouds here, are unconfined i.e. there are no ICM particles. Fig. 5 below shows a time sequence of column density plots of the colliding clouds through to the development of the NTSI in the shocked slab. The initial behaviour of the post-collision gas slab is similar to that in the previous case, eventually the NTSI  dominates the dynamical evolution of the slab. The gravitational instability thus, seems to be suppressed. Another instability to which the slab is susceptible, is the Kelvin-Helmholtz (KH) instability. The distinct feature of which is, the formation of vortices. Material from colliding clouds streams in from either sides of the contact layer, and smears laterally after striking it.  Obviously, the gas elements within the slab are in a turbulent state and artificial viscosity is responsible for shearing interaction between them.

Sure, SPH in its simplest form is unable to intricately resolve the vortex rolls associated with the KH instability but it can at least represent the underlying physical process, that of layers mixing. That, this produces a non-uniform velocity field in the plane of the slab is evident from the plot in the top panel of Fig. 6. This is a 3-d plot showing the  velocity field surface in the plane of the slab. Number of secondary peaks on this surface correspond to the localised circular velocity patterns in the slab, obvious features of the KH instability.  In the bottom panel, the velocity field has been overlayed on the column density plot of the slab as seen by an observer stationed in a direction orthogonal to the plane of collision. From this plot we can see that the slab develops kinks and the bending modes are more readily visible in its central plane.

With the help of first order perturbative  analysis, it can also be shown that fluid surfaces in shearing contact develop surface gravity waves which give them a rippled appearance also called, the breathing modes of the fluid surface. This situation is analogous to the formation of ripples on the surface of water bodies (Raichaudhari 1999; Shore 2007). Fig. 7 below is a column density plot showing a face-on view of the shocked slab. Density contours have been overlayed on it to facilitate identification of structure within the slab. We can see a dense  network of filamentary structure and clump ($\rho\sim 10^{-18}$ g cm$^{-3}$) formation is visible in regions where filaments intersect. However, this structure suffers from tidal disruption due to strong shearing motion between slab layers. Klessen \& Burkert (2000) and Vietri, Ferrara \& Miniati (1997) for instance, make similar observations about ablation of clumps in their models of turbulent GMC interiors.

 Since we start with unconfined clouds, the shocks confining the gas slab move outwards (essentially vacuum in the simulation) and the slab re-expands and then collapses laterally. See the plot of particle positions corresponding to $t$ = 0.50 Myr in Fig. 12 below. The thermal pressure within the slab is insufficient to arrest the collapse, which proceeds towards the formation of a filament along the  collision axis. Gravitational clustering may then commence in this filament. However, we have not followed the simulation to that point due to constraints on resources.  We note that grid codes simulating colliding clouds  suggest that the re-expansion followed by the collapse phase, is dominated by the Rayleigh-Taylor (RT) and KH instabilities (c. f. Miniati \emph{et al.} 1997). However, there is no evidence for the RT instability in this model, although vortex formation in the slab suggests  existence of the latter.

\begin{figure*}
\vbox to 180mm{\vfil 
\includegraphics[angle=270, width=14.0cm]{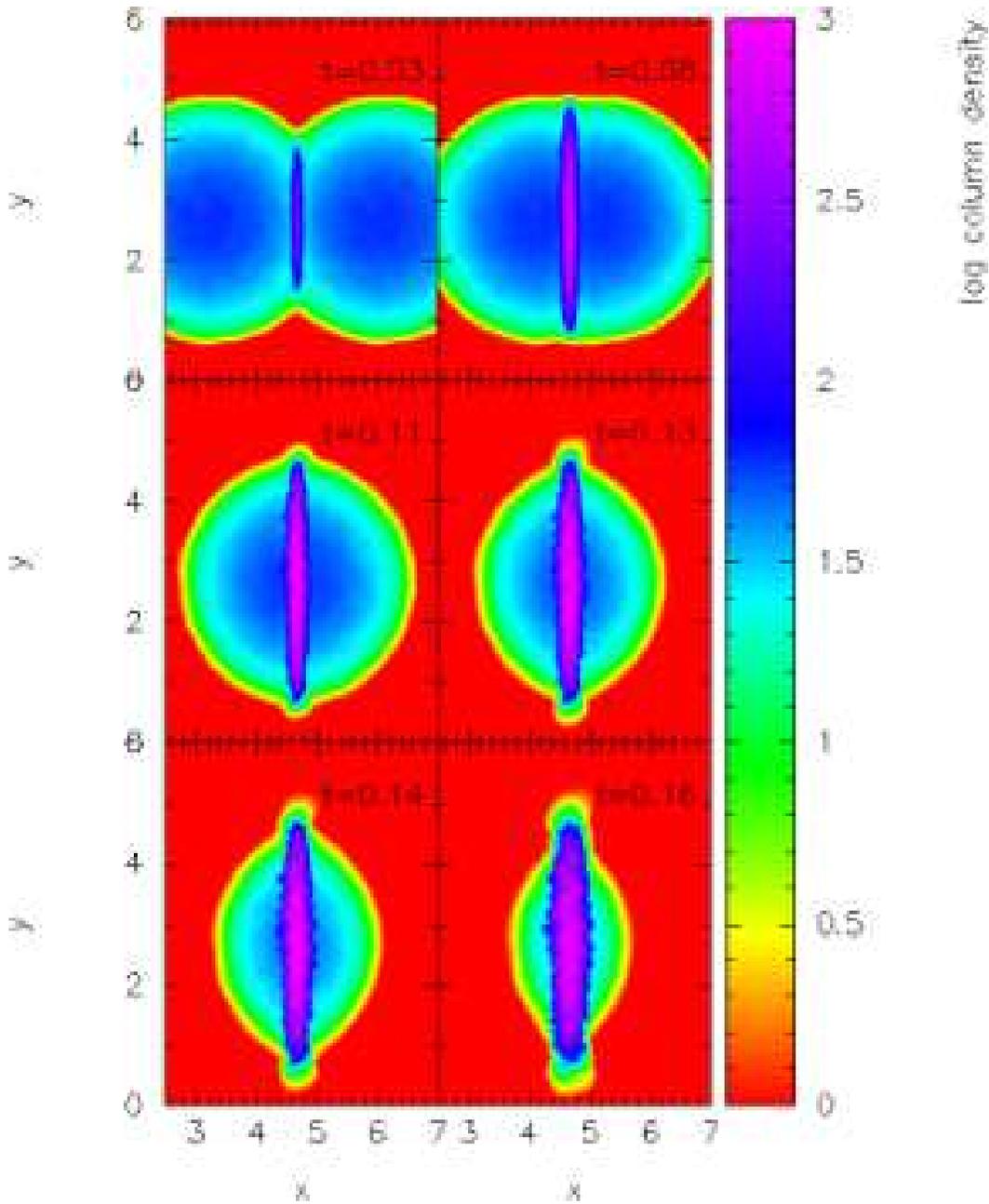}
\caption{A time (measured in Myr) sequence of the column density plots in model 3. Formation of the shocked slab can be seen in the snapshot corresponding to  $t$ = 0.03 Myr. The NTSI then dominates the slab, as can be seen in the snapshots in the second and third rows, after which the slab soon becomes bloated ($t$ = 0.16 Myr).} 
\vfil} \label{landfig} 
\end{figure*}

\subsubsection{Model 4}

 In this case unconfined  clouds collide at a finite impact parameter, $b = \frac{R_{cld}}{4}$, with other physical parameters the same as those in model 2. The supersonic cloud collision results in an oblique shock compressed slab. 

\begin{figure*}
\centering
  \vbox to 240mm{\vfil
        \includegraphics[angle=270, width=9.5cm]{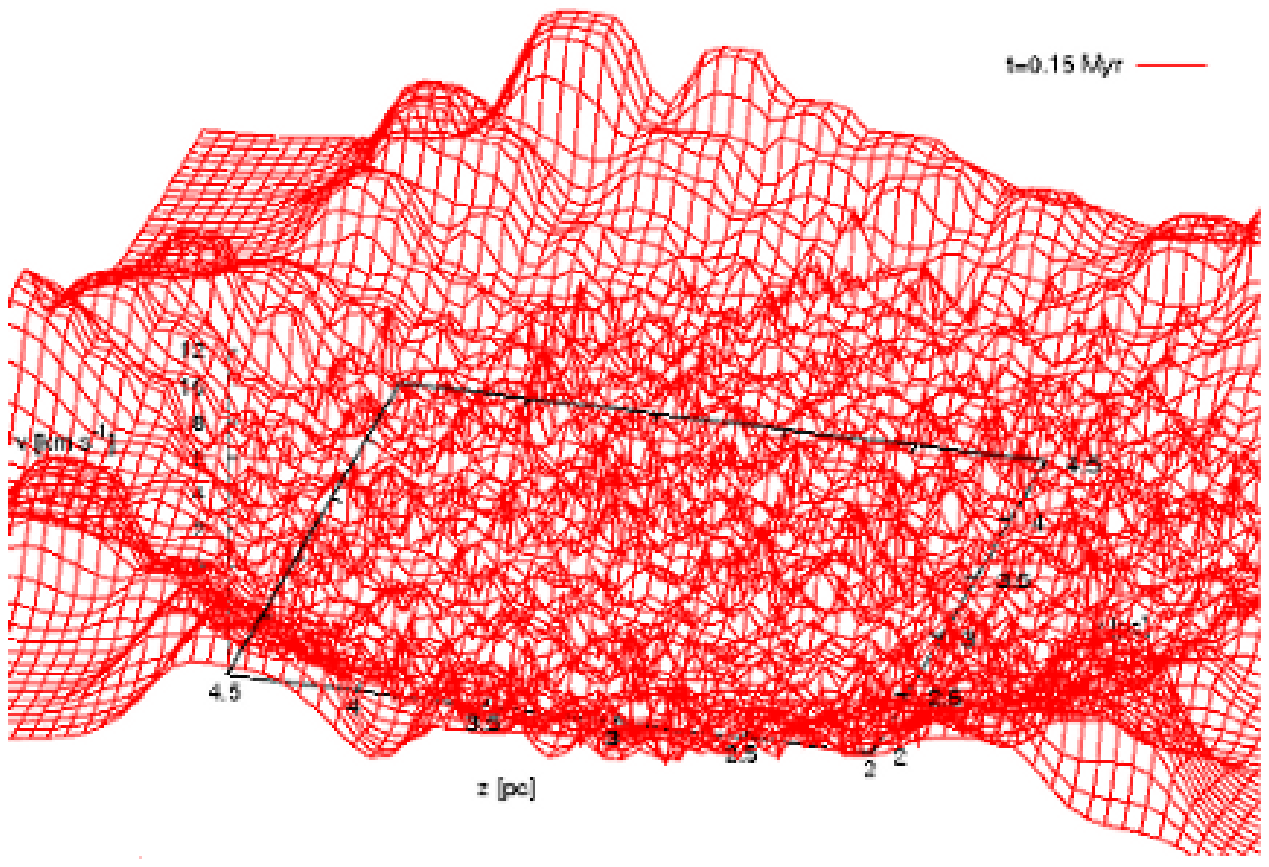}
         \includegraphics[angle=270, width=9.5cm]{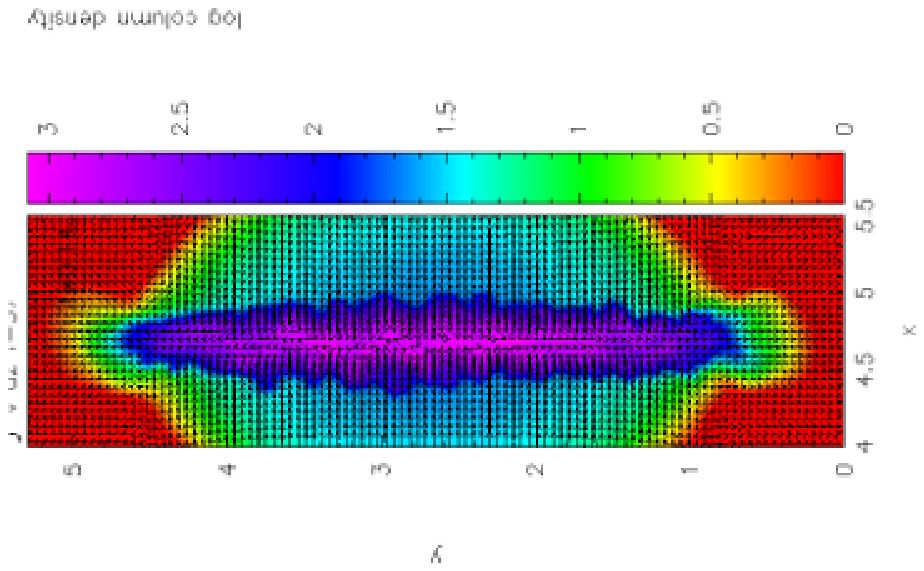}
  \caption{\emph{Top panel} : A plot of the velocity field ($t$ = 0.14 Myr) in the plane of the slab in model 3. It is evident that the velocity field within the slab is highly uneven and the spikes in it suggest formation of vortices in the slab.  \emph{Bottom panel} : Velocity field in the plane of collision overlayed on column density plot of the slab ($t$ = 0.14 Myr) in model 3. The kinks in the slab along with the \emph{jets} are evident from this plot. Velocity is measured in km s$^{-1}$. } \vfil}\label{landfig}
\end{figure*}

\begin{figure*}
\vbox to 110mm{\vfil 
\includegraphics[angle=270, width=13.0cm]{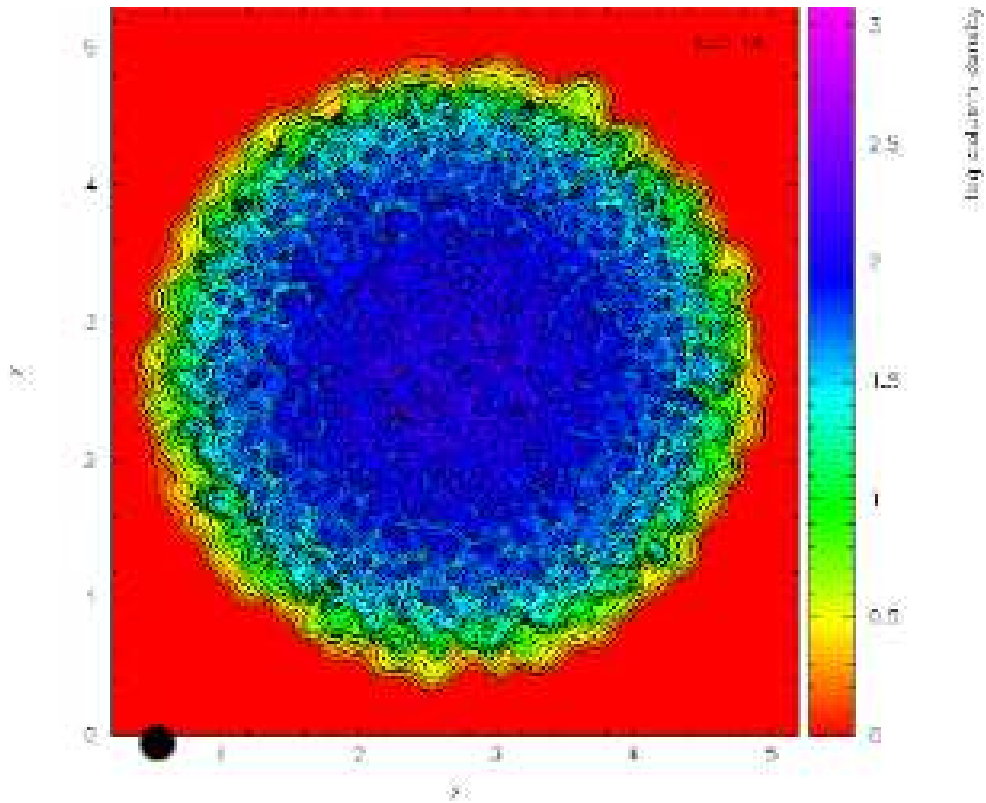}
\caption{A column density plot of the shocked slab in model 3, as seen in the plane orthogonal to the plane of collision. Density contours overlayed on it show several elongated features and dense clumps within them. Time measured in Myr.} \vfil} \label{landfig}
\end{figure*}

 \begin{figure*}  
\centering
\vbox to 130mm{\vfil
\includegraphics[angle=270, width=13.0cm]{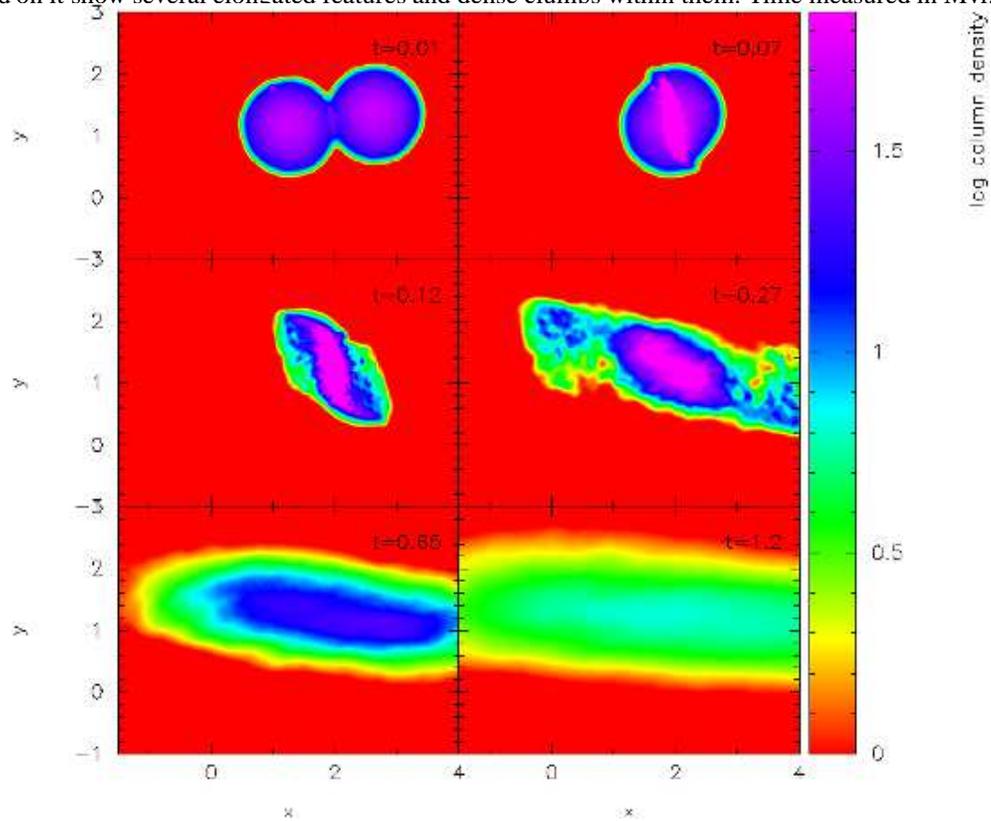}
\caption{A time (measured in Myr) sequence of column density plots in model 4. The off-centre cloud collision leads to the formation of an oblique shocked slab as can be seen in the pictures of the top row. In the second row we can see that the slab tumbles about the $z$ axis and breaks into three parts, with the outer two lobes moving away from the central region. In the bottom row we see that the central region evolves into a filament aligned with the collision axis.    } \vfil}\label{landfig}
\end{figure*}

\begin{figure*}
\centering
\vbox to 110mm{\vfil 
\includegraphics[angle=270, width=10.0cm]{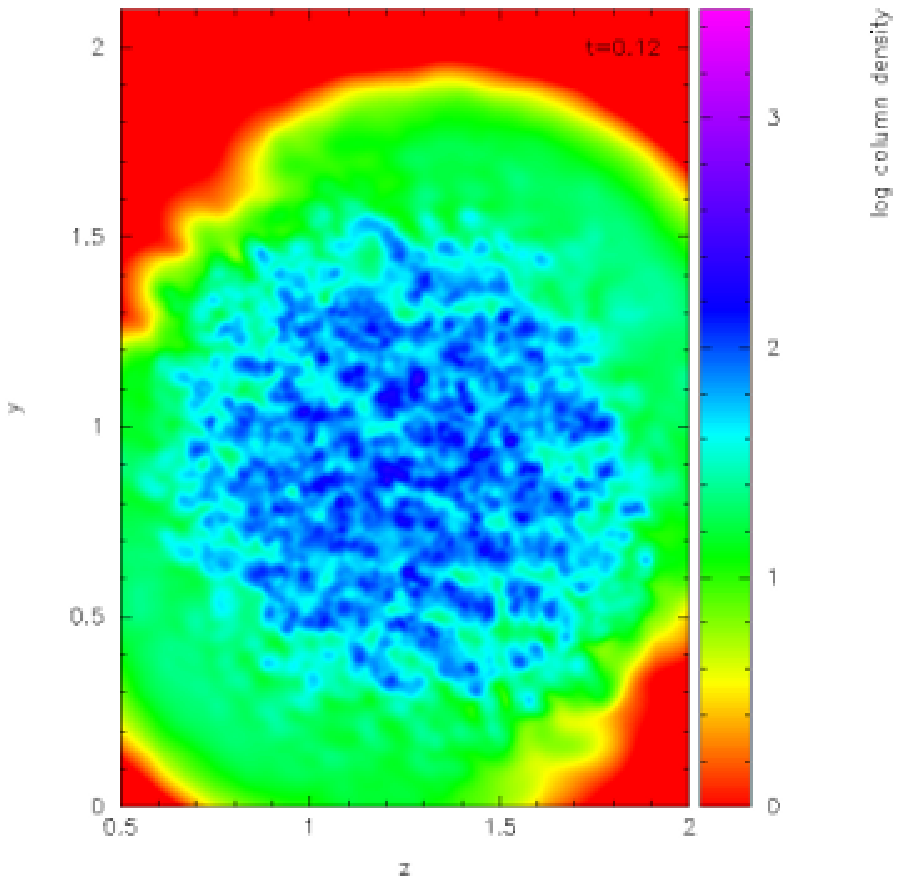}
\caption{A column density plot of the shocked slab in model 4, shortly after its formation. Structure within the slab is evident from the plot along with dense clumps, at the intersection of filaments.  Time measured in Myr. } \vfil} \label{landfig}
\end{figure*}

Unlike the planar slabs formed in a head-on collision, the oblique shocked slab  has three distinct regions. The central planar region and the two arched extensions at the top and bottom ends of the slab. See snapshots in the panels of Fig. 8. These extended features are due to the fact that the initial shock front effectively blocks the fluid flow from either sides in the central regions, while at the top end where the fluid  flows from the right to left, has higher transverse velocity compared to that in  the central regions. For the same reason there is also an extension at the lower end of the slab, except that the fluid at this end is flowing from the left  to right. 

Like the former cases, the shocked slab in this case is also internally stressed and evolves into a  filament aligned with the collision axis. Like the previous two cases, the shocked slab in this case also quickly develops structure in it. Elongated density structures and blobs can be seen in Fig. 9 below, which is a column density plot of the slab as seen face-on.

\section{Discussion}

Collision between clouds is characterised by three phases - (\emph{a}) \emph{the compression phase}, during which post-collision, a contact layer  forms first and shock waves propagate in individual clouds. The slab is confined by ram pressure on its either side which strongly  compresses the gas within the slab. The thickness of the slab depends on the effectiveness of the artificial viscosity. Additionally, it could also depend on the efficiency of radiative cooling, but such is not the case in the present work as we use a plain EOS. 

(\emph{b}) The next phase is that of \emph{re-expansion}. During this phase the shocks confining the slab propagate outward in the ICM and a rarefaction wave propagates back into the slab. This creates a central low pressure and low density region in the slab. During this phase, some material squirts out from the top and bottom of the slab. This is also the phase during which the slab may become susceptible to the NTSI and the KH instability due to lateral motion of the gas within the slab at supersonic speed (relative to the local sound speed in the slab). 

(\emph{c}) The final phase is that of lateral collapse. The slab expansion is halted by the  ICM by which time, pressure within the slab becomes much smaller than the external pressure. Consequently the expanding slab material is driven towards the central low density region. 

(\emph{d}) If the externally confining pressure is too weak, the final phase could be different and the re-expanding slab may simply form a diffuse cloud, amalgamated with the ICM.

\subsection{Instabilities in a shocked slab}

The problem under investigation here is that of the stability of a viscous fluid slab confined by accretion shocks on its either sides.  Growth of instabilities in a cold, isothermal slab has been extensively studied in the past. Linear perturbative analyses have shown the existence of various unstable modes, see for instance Ledoux (1951), Chandrasekhar (1953), Mestesl (1965) and Simon (1965). However, these investigations assumed slabs with no external forces acting on them.

A shocked slab on the other hand, shows  propensity for a variety of instabilities. However, the present investigation is limited to the gravitational and shearing instabilities only since other relevant physical details have been avoided in this work. The final state of the shocked slab depends on the relative dominance of either of the two instabilities. The shearing instabilities are mainly of two varieties- (a) the non-linear thin shell instability (NTSI) (b) the Kelvin-Helmholtz (KH) instability.  The condition for the dominance of the gravitational instability over the shearing instabilities, especially the NTSI, is given by equation (13) above. \\
 \subsubsection{The NTSI}

Growth of instabilities in shocked slabs was first studied by Vishniac (1983) by invoking full shock conditions, who suggested the existence of bending modes  of a shocked slab. His linear analysis of the growth of these modes showed that the instability became overstable and led to disintegration of the slab.  The resulting fragments suffered erosion due to the RT instability. Gravitational instabilities are suppressed in this regime. In the limit of a small bending angle, the linear analysis is sufficient to predict the growth rate of the unstable mode. The linear theory however, breaks down when velocities within the slab become comparable to the local sound speed and fluctuations in column density become of order unity.

This investigation was extended by Vishniac (1994) to the non-linear regime of the  bending and breathing modes. The analysis showed that  an isothermal shocked slab of inviscid fluid,  became unstable due to non-linear growth of the unstable modes.   Since this instability commences in thin shells (i.e. between layers of the slab), it is called the Thin Shell Instability. The instability grows rapidly and saturates leading to turbulent mixing between layers of the slab. The slab eventually forms a sieve like structure and becomes bloated before finally disintegrating to form a filament aligned along the collision axis. Gravitational clustering can then occur in this filament. A similar observation has also been reported by Hueckstaedt (2003) in grid simulations of colliding fluid flows. According to Burkert \& Hartmann (2004) a cold, self gravitating gas slab collapses over a timescale, $t_{c}$, and
\begin{displaymath}
t_{c} = \Big(\frac{R_{cld}}{\pi G\Sigma_{cld}}\Big)^{\frac{1}{2}},
\end{displaymath} 
where $\Sigma_{cld}$ is the slab surface density. In case of model 3 for instance, $t_{c}$ is of order 1 Myr and from Fig. 12 below, we can see that the gas slab has collapsed substantially by $t$ = 0.5 Myr.

Although we start with unperturbed initial conditions, the perturbations that develop in the shocked slab are purely from white noise. However, the simulation has sufficient resolution to resolve the observed NTSI wavelength, $\lambda_{NTSI}$. In SPH, the spatial extent of the smallest region resolved is of the order of the particle smoothing length, $h$. An SPH sphere of influence has volume $32\pi h^{3}/3$, so that for a shocked slab with $N_{gas}$ particles,
\begin{displaymath}
h = \Big(\frac{3R^{2}_{cld}N_{neibs}\triangle x}{32 N_{gas}}\Big)^{\frac{1}{3}},
\end{displaymath}
where $\triangle x$ is the shock thickness and remaining symbols have their usual meanings.  For the chosen dimension of clouds, the smoothing length is at least an order of magnitude smaller than $\lambda_{NTSI}$. We therefore believe that the features of the instability observed in our simulations are physical.

 As a shocked slab accretes material, small perturbations i.e. the bending modes of the slab get amplified due to non-linear transfer of momentum within the slab. Material streaming in the perturbed gas slab deposits a net positive momentum in the convex regions and a net negative momentum in the concave regions of the slab.  Transfer of momentum within the slab amplifies its bending modes to scales greater than the thickness of the slab itself. It is essential to note that the pressure within the slab is thermal and isotropic while the external one, is ram pressure. This distinction is irrelevant in the unperturbed state.

However, once a bending mode of the slab is excited, there is a net imbalance in forces acting on the convex and concave regions of the slab leading to its amplification.  In general, the accreted material converges in the convex regions of the slab and 
 diverges in the concave regions. Slab layers undergo turbulent mixing  due to strong shear between them. If $\tau$ is the dynamic time scale of the shocked slab (i.e. the time scale on which physical properties of the slab evolve dynamically), turbulent mixing occurs when $\tau$ is greater than the time required for  generation of the next similar layer at a distance greater than the wavelength and having wavenumber $k$. If $L$ is the thickness of the slab and $t$ is its age then
\begin{equation}
\tau\gg\frac{t}{kL},
\end{equation}
 is the condition under which turbulent mixing is suppressed Vishniac (1994).

If $\tau\sim t$, the turbulent mixing is weak and is only marginally effective in the post-shock slab. The turbulent mixing of slab layers gives rise to eddies on the scale of slab thickness which is an effective means of energy dissipation. The NTSI grows at a rate,
\begin{equation}
\tau^{-1} \sim a_{0}k(kx_{m})^{1/2}\cdot C^{-1/2}_{d}
\end{equation}
Vishniac (1994).

Here $x_{m}$ denotes the mid-plane of the slab and $C_{d}$ is a constant of order unity. This constant describes the extent of dissipation resulting from mixing  between slab layers.  Klein \& Woods (1998) estimate $2\lesssim C_{d}\lesssim 4$ and a rate of saturation $>a_{0}k(kL)^{1/2}$. 

\subsection{Observations from simulations}

Clouds collide and first form a layer of contact after which, shock waves propagate into individual clouds. The cloud material streams into the layer and the timescale on which a cloud is destroyed is called the cloud crushing time, $t_{cr}$. Shock waves  propagating in a cloud reach its far edge in this much time, and $t_{cr}\sim\frac{2R_{cld}}{v_{s}}$ (Klein \& Woods 1998). Here $v_{s}$ is the velocity of the shock wave.

The density and pressure (thermal + ram) gradients respectively,  have opposite signs across the shocked slab. The slab surfaces are shocked while pressure within the slab is thermal. Thus the pressure gradient across the slab surface has opposite sign relative to  the density gradient, that increases steeply. The shocks confining the slab apparently arrest the inflow of gas from the individual clouds. The colliding clouds  are therefore held up. 

This can be seen for instance, in model 2 (\S 3.3.2). The column density plot corresponding to $t$ = 0.085 Myr (which is greater than $t_{cr}$) in Fig. 3 above, shows that the colliding clouds survive even beyond their crushing time, $t_{cr}$.  In fact at this epoch, the shocked slab has undergone gravitational fragmentation. Fig. 4 shows the column density plot ($t$ = 0.085 Myr) of the fragmented slab. Filamentary structure within the slab is evident in this plot. The filaments then develop condensations and form a number of clumps in them. These clumps can be seen as bright pink spots in the picture. 

The clumps are actually sink particles and we remind our reader that sinks, in the present work represent protostellar cores. Soon after its commencement, core formation spreads like an epidemic and within a few thousand years, sixteen cores are formed in various filaments, whence the simulation was terminated. Cores accrete material channelled along their natal filaments.

Immediately after its formation, the shocked slab develops kinks and features of the NTSI are visible in it. The bending modes of the slab subsequently amplify and soon, it becomes bloated. This can seen in the snapshots corresponding to $t$ = 0.03 Myr, $t$ = 0.05 Myr and $t$ = 0.07 Myr of Fig. 3. Interestingly, there is a confluence of the gravitational instability and the NTSI but the former dominates over the latter and the condition for dominance  of the gravitational instability is given by equation (13) above. Thus, we note that the gravitational instability can dominate the shocked slab even in presence of the NTSI, provided the slab is sufficiently dense. 

Although we have seen evidence for clump formation in this model, only the self gravitating clumps survive. Other smaller clumps either merge to form larger clumps or simply diffuse away. \emph{We therefore envisage this model as one leading to the formation of filamentary structure, that may further spawn star formation}. This is therefore an interesting case from the perspective of star formation. However, we do not observe any evidence for secondary fragmentation of cores in this model.

In model 3 we perform a similar experiment with higher resolution but no intercloud medium.  The shocks confining the post-collision slab move outward (into vacuum) and thus, permit re-expansion of the slab, as can be seen in the pressure-distance plots of Fig. 11. Distance in parsecs is marked on the $x$ - axis and pressure, normalised relative to the initial pressure, has been marked along the $y$ - axis. The plots in the top row ($t$ = 0.48 t$_{cr}$), correspond to the post collision jump in density and pressure, respectively. The plot of pressure shows the formation of shocks at the slab surfaces while that within the slab is thermal and uniform.

The pressure and density gradients respectively, have opposite signs across the slab surfaces, making it unstable. Unlike  Hunter \emph{et al} (1986), we are however, disinclined to term this situation as Rayleigh-Taylor (RT) unstable. The outwardly propagated and weakened shocks can be seen in the plots corresponding to $t$ = 11.4 $t_{cr}$. During this phase the central density and pressure within the slab falls, as can be seen in the plots in the second and third row of this figure.  Spikes in the density distribution indicate clump formation which obviously, do not survive. The slab continues to re-expand and flattens laterally. Eventually it becomes elongated with a large aspect ratio and its long axis lying in the plane of collision. Fig. 12 shows a plot of particle positions in the plane of collision.

As the shocked slab evolves, a few breakaway blobs are visible near its surface. These can be seen in the column density snapshot corresponding to $t$ = 0.16 Myr in Fig. 5. This is perhaps due to the inability of  SPH in its simplest form  to model the KH instability, apparently due to formation of artificial boundaries between regions having steep density contrast (Agertz \emph{et al} 2007). This problem can be fixed by including some additional correction terms to account for  thermal energy dissipation across these boundaries (Price 2007 II). Our code does not include these corrections, however this shortcoming does not seem to have any adverse bearing  on the results of the simulations and in any case it is not our aim here, to model the KH instabilities in their minute details.

 Similar breakaway blobs, having much smaller densities (at least by two orders of magnitude), as compared to those observed in our simulations, have been reported by Heitsch \emph{et al} (2008). They interpret this observation as evidence for core formation. While these latter authors began with perturbations on lateral faces of colliding fluid streams and grid codes apparently, being able to better handle density contrasts, it is plausible that their observation of core formation is physical however, we prefer to discount the blobs as numerical artifacts despite they having densities about an order of magnitude larger than that of the slab.

Further, as in the previous case, the shocked slab in this case also shows features of the NTSI. See column density plots corresponding to $t$ = 0.11 through to 0.16 Myr, in Fig. 5 above. These pictures show that the kinks (bending modes) in the slab, grow rapidly and soon it becomes bloated ($t$ = 0.16 Myr). In the bottom   panel of Fig. 6, the local velocity field has been overlayed on the column density plot of the shocked slab ($t$ = 0.14 Myr), by which time features of the NTSI have developed well.  This picture shows amplified kinks in the slab. By measuring the amplitude of the perturbation and the slab thickness, we calculated the growth rate of the NTSI as
\begin{equation}
\tau^{-1} \sim (0.56)(kL)^{0.78},
\end{equation}
where $k$ is the wavenumber of the unstable mode and $k = \frac{2\pi}{\lambda_{NTSI}}$, $L$ is the thickness of the slab.  Fig. 10 shows the growth rate of the NTSI. The measured growth rate has been plotted in red while the smooth green curve is a power-law fit (equation 16) to the one in red.

The observed growth rate of the NTSI in this simulation is in good agreement with that reported by  Blondin \& Marks (1996) and Klein \& Woods (1998). They  suggest a growth rate of order $(kL)^{0.7}$. These respective growth rates of the NTSI however,  do not match with that proposed by Vishniac (1994) and given by equation (15) above. This later equation predicts a much faster growth ($k^{1.5}$), of the instability. In fact the instability in this case, has been observed to grow on a timescale comparable to the sound crossing time across the shocked slab.

We have argued that gas within the shocked slab is in turbulent motion, so that the slab is internally shocked and mechanical energy dissipated. Accompanied with this, is the loss of linear momentum of gas elements within the slab. The larger the viscosity, the greater is the dissipation. We have already noted that growth of the NTSI is related to momentum transfer between perturbed regions of the slab. This dissipation should heat the gas up, a phenomenon that is obviated due to a barytropic EOS.  This is followed by lateral expansion of the gas slab, which soon becomes bloated. Rounding of the perturbed surfaces will reduce local pressure imbalances, another crucial factor inconducive to the growth of NTSI. Concurrence of these two factors may be responsible for arresting the growth of this instability.  The NTSI is observed in models 2 and 3 and grows at a comparable rate in either cases. Artificial viscosity being the only common factor between the two cases, is therefore likely to have influenced its growth. Other cited literature here however, does not discuss the possible reasons for the observed slow growth rate. \emph{Looking at the end product of this simulation, it may be suggested that such a collision is likely to produce a diffuse cloud which may collapse further and form a long filament. Gravitational clustering may commence in this filament}. 

\begin{figure}
        \includegraphics[angle=270, width=7.50cm]{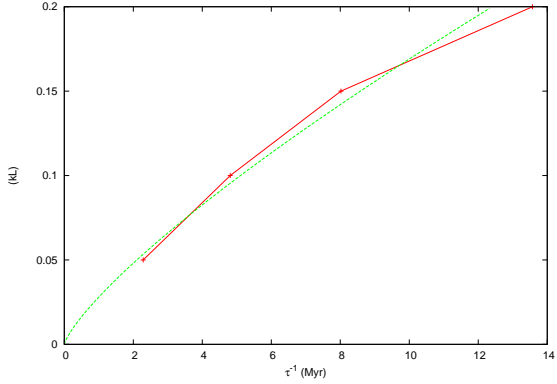}
  \caption{The growth rate of the NTSI in model 3. The red curve corresponds to the actual growth rate of NTSI in the simulation while the green curve is a power law fit (Eqn. (16)) to the observed growth rate. Time on the $x$ axis is measured in units of the cloud crushing time $t_{cr}$.}
\end{figure}

\begin{figure*}
\vbox to 180mm{\vfil 
        \includegraphics[angle=270, width=22.0 cm]{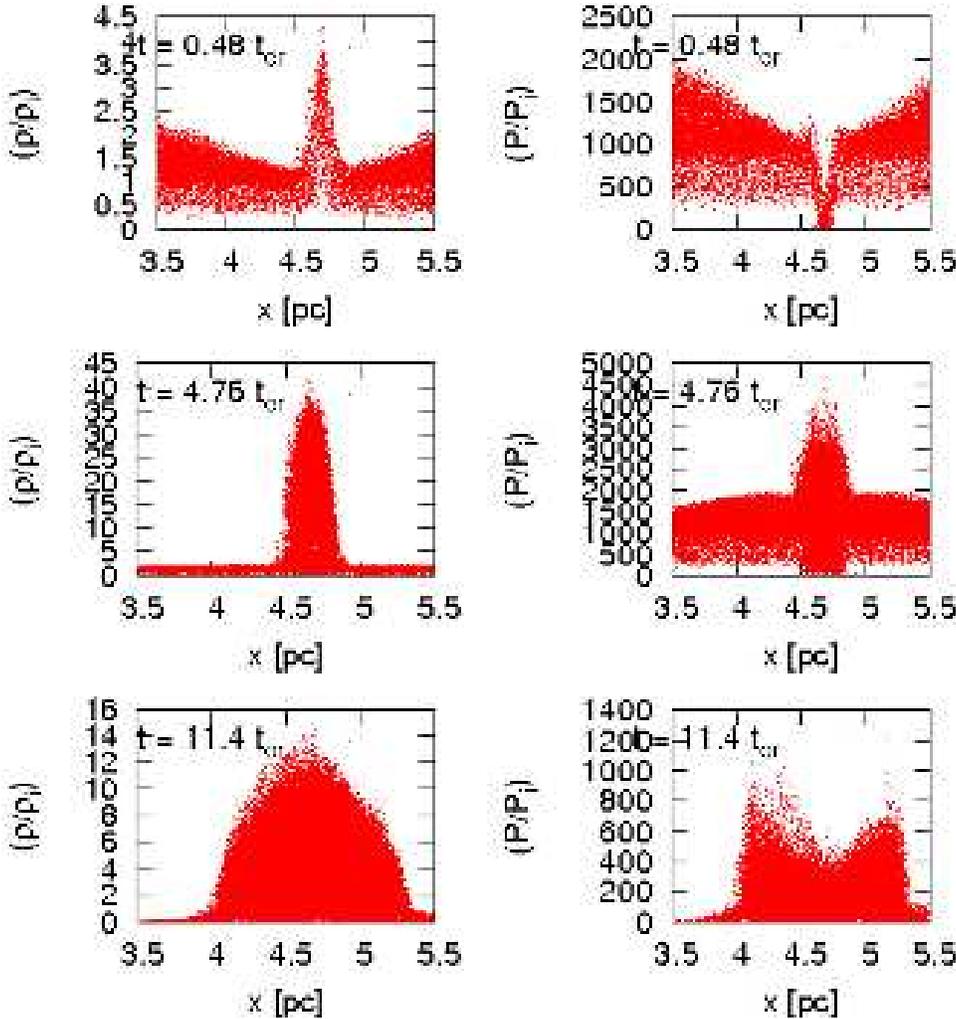}
  \caption{Time evolution of density and pressure against distance along the collision axis in model 3. Time is measured in units of the cloud crushing time, $t_{cr}$. Both, density and pressure have been normalised relative to their respective initial values. In the top panel we can see the jump in density and pressure, immediately after cloud collision. Note that the average density contrast behind the shock is roughly four, as should be the case for a strong adiabatic shock. The density behind the shock increases further, as we move in to the isothermal shock approximation, see Eqn. 10 above. The shocks then move outward and the slab begins to re-expand, accompanied with a drop in density and pressure within the slab. We can see this in respectively, the second and third panels above.}\vfil} \label{landfig}
\end{figure*}

\subsubsection{Off-centre cloud collision}

The shocked slab is internally stressed  and thus exhibits a non-uniform velocity field. Internal shear renders the slab layers KH unstable. The picture in Fig. 9 is a column density plot of the slab soon after its formation. The slab forms well defined structure which is either filamentary or globular. Some small globules merge with other similar globules to form larger ones while a few others are destroyed.  As in model 3, the colliding clouds and therefore the post-collision gas slab, are unconfined. The confining shocks can therefore propagate outward. Thus, the slab in this case as well, passes through the same phases as the one in the previous case. Additionally, it also  exhibits rotational motion about an axis orthogonal to the plane of collision ($z$ axis).  As the slab tumbles, two satellite lobes break off and move away radially. See the snapshot corresponding to $t$ = 0.27 Myr in Fig. 8.

 The slab dissipates thermal support through internal shearing motion. Usami, Hanawa \& Fujimoto (1995) provide a detailed analytic account of oblique shock compressed slabs, according to which gravitational instabilities are suppressed in such slabs. Shearing motion increases the effective sound speed and therefore damps the growth rate of gravitational instability, which is probably exacerbated by the type of EOS employed in this work.

Further, these authors also predict the end product of such slabs to be an elongated object, similar to the one we observe at the end of this calculation. We terminated this calculation after the formation of a filament in the plane of collision. The sequence of colliding clouds and then formation of the filament has been shown in a time sequence of column density plots in Fig. 8 above. We note that the slab in this case evolves on a much longer timescale as compared to the earlier models. This is presumably due to the additional support provided by the angular momentum associated with the slab rotation.

We notice some perturbations on the slab surface (see snapshots for $t$ = 0.07 Myr and $t$ = 0.12 Myr in Fig. 8) but do not amplify as much as those in the earlier model. The NTSI thus, seems to be damped out.  The Coriolis force acting on the rotating slab may suppress perturbations normal to the slab surface, in other words, the bending modes of the slab. This is however, contrary to the findings of Whitworth \emph{et al} (1995) who simulated off-centre low velocity cloud collision and found that the pressure compressed oblique slab, underwent gravitational fragmentation. However, with just a few thousand particles it is unlikely that they would have been able to resolve the gravitational instability.

\section{Conclusions}

  Cloud collision plays an important role in the interstellar gas dynamics and  is one of the mechanisms of dissipating energy.  Sufficiently dense clumps resulting from fragmentation of the post collision gas slabs become self gravitating and spawn single/multiple star formation. A violent phenomenon like cloud collision also replenishes the pool of elements in the galactic disk.  Post shock gas attains incredibly high temperatures ($\gtrsim 10^{4}$ K) however, if the shocked gas radiates efficiently, it cools rapidly and the temperature soon drops, sometimes even to the initial preshock temperature. Gas is further compressed by the shock and postshock gas bodies may attain densities as high as $\sim10^{4}\textrm{cm}^{-3} - 10^{5}$ cm$^{-3}$. Under favourable conditions, star formation may commence in these gas bodies . Thus, gas in the ISM goes through a cycle of phase transition starting from- diffuse gas to MCs  to shocked slabs to star forming cores to stars and back to diffuse gas. In some cases however, stars just may not form between the shock phase and the final diffused state. 

The scope of the cloud collision problem is quite large and here we attempt to investigate only its dynamical features. With the aid of the simple models tested in this work, we have attempted to suggest that cloud collision provides a viable mechanism for the formation of elongated structure in the ISM. The densest regions in these filaments may become self gravitating. We have  concentrated only on the gravothermal aspect of the matter, thereby discounting  the effect of the interstellar chemistry. We have tried to ascertain the role of dynamical instabilities on the evolution of shocked gas slabs and have shown that, there is much more to the problem than just a balance between gravity and thermal pressure. Fig. 13 below is a plot of the normalised gravitational and thermal energy against time, for models 2, 3 and 4. A common feature of these plots is that, post-shock, the thermal energy rises steeply, attains a maximum before rapidly falling off during the re-expansion phase of the slab. However, the shocked slab in model 2 expands only marginally,  and undergoes gravitational fragmentation. The slab in this case is thermally supported, as is evident from the red curve in Fig. 13. The respective slabs in models 3 and 4, re-expand considerably and eventually end up at a temperature close to that in the precollision clouds. The gravitational energy, in models 3 and 4, is almost in sync with the thermal energy; with model 2 being an exception. The cold shocked slab in this latter case becomes gravitationally unstable and fragments. The steep rise in gravitational energy corresponds to self gravitating fragments.

In all the models tested here, the shocked slab becomes bloated soon after its formation which is a unique feature of the NTSI.  The NTSI along with the KH instability, play a vital role in energy dissipation via vortex formation within slab layers. Evidence for the NTSI has also been observed in grid simulations of colliding turbulent fluid flows (Heitsch \emph{et al} 2008). However, the gravitational instability dominates a shocked slab provided it is sufficiently dense else it is suppressed and the slab re-expands. This can be seen in models 1, 3 and 4. Usami, Hanawa \& Fujimoto (1995) for instance, make a similar  claim. The NTSI appears to dominate the unconfined slab in model 3, while model 1 is poorly resolved, although the final result is the same. The slab in model 2, contained by the ICM, is more interesting as we observe the NTSI competing with the gravitational instability and finally the slab fragments. The oblique shocked slab in model 4 develops kinks which seem to be damped and so, nothing conclusive can be said about growth of the NTSI in this model.

Klein \& Woods (1998) and Heitsch \emph{et al} (2006)  also report similar conclusions. The former additionally note that, the NTSI is suppressed in adiabatic shocked slabs as  such shocks do not  radiate post-shock kinetic energy and therefore tend to rapidly re-expand, on a timescale much shorter than that required for the growth of bending modes. We measure the growth rate of the NTSI in one of our models, which agrees very well with that reported in literature elsewhere.

In the present work, we have  adopted a simple EOS to account for radiative cooling and the assumption of an isothermal shock may have led to excessive post-shock compression. It is therefore essential to employ better cooling techniques (e.g. Koyama \& Inutsuka 2000; V{\`a}zquez-Semadini et al 2007). It is also essential to investigate the effect of SPH artificial viscosity on the growth rate of NTSI. We adopted the standard choice of viscous parameters in all our simulations here. The models presented here are extremely simple, since we have preferred to use physically identical  clouds without any internal perturbations. Also, we have not included magnetic fields in the present work. It would be an instructive exercise to start with magnetised clouds and then collide them. We can then study the effect of  magnetic field on the evolution of the shocked slab and in particular, on the growth of various dynamical instabilities in a magnetised gas slab.

\begin{figure}
\centering
        \includegraphics[angle=270, width=8.0cm]{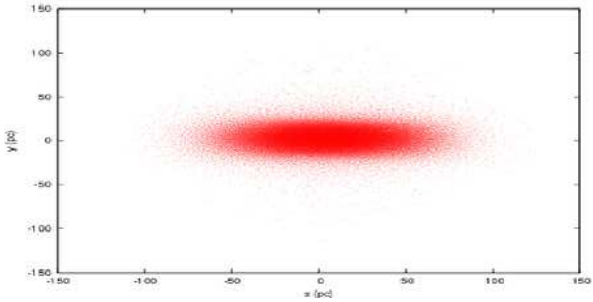}       
  \caption{A plot of particle positions showing the formation of an elongated object in model 3. At this epoch ($t$ = 0.5 Myr), the shocked slab has re-expanded and undergone significant flattening.}
\end{figure}

\begin{figure}
\centering
        \includegraphics[angle=270, width=7.0cm]{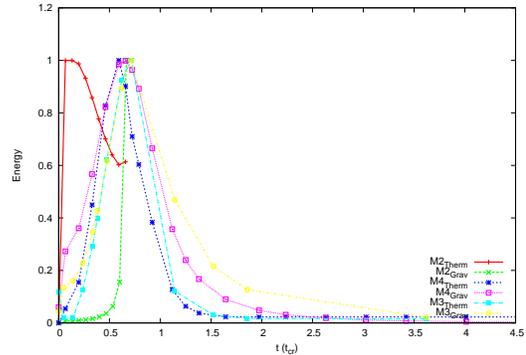}       
  \caption{A plot showing the time evolution of the normalised hydrodynamic and gravitational pressure in models 2, 3 and 4. Time along the $x$ axis is marked in units of $t_{cr}$. This plot suggests that the shocked slab is supported against gravity unto the re-expansion phase, after which gravity starts dominating and the slab collapses. The key in the right hand corner lists the markers (M$_{energy}$) used for respective models.}
\end{figure}

\begin{acknowledgements}
     This work was completed as part of post-graduate research and funded by the State Government Of Maharashtra, India (DSW/Edu/Inf/2004/6283(35)). I thank Dr. S. Goodwin for making available the latest version of his SPH code DRAGON. All the column density plots presented here have been prepared using the publicly available graphics package, SPLASH prepared by Dr. D. Price (Price 2007I). Useful comments and suggestions by an anonymous referee are greatly appreciated and special thanks to Prof. S. Falle for his critical observations of the original paper, which contributed immensely towards its improvement.

\end{acknowledgements}

\end{document}